\documentclass[prd,superscriptaddress,twocolumn,showpacs,nofootinbib]{revtex4}
\usepackage{graphicx}
\usepackage{amsfonts,latexsym}
\usepackage{amsmath,amssymb}
\usepackage{graphicx,color}
\usepackage{multirow}
\newcommand{\nn}{\nonumber}
\newcommand{\be}{\begin{equation}}
\newcommand{\ee}{\end{equation}}
\begin{document}
\input epsf
\title{ Hairy  mass bound in the Einstein-Born-Infeld black
hole}

\author{Yun Soo Myung}\email{ysmyung@inje.ac.kr}
\affiliation{Institute of Basic Sciences and School of Computer
Aided Science, Inje University Gimhae 621-749, Korea}

\author{Taeyoon Moon}\email{tymoon@sogang.ac.kr}
\affiliation{Center for Quantum Space-time, Sogang University,
Seoul, 121-742, Korea}

\begin{abstract}
We study the Einstein-Born-Infeld (EBI) theory where the hairy black
hole was found as the EBI black hole by using Hod's idea. The hair
extends beyond the photon-sphere of the EBI black hole spacetimes.
We show that the region beyond the photon-sphere involves more than
half of the total hair's mass, which respects Hod's conjectured
bound.
\end{abstract}
\pacs{04.70.Bw, 04.60.Kz, 04.50.Kd} \maketitle It is known that when
a black hole is not completely determined by the global three
charges of ADM mass $M$, electric charge $Q$, and angular momentum
$J$ defined at infinity, but rather it possesses short range charges
(hair) that vanish at infinity, one calls it a hairy black hole. It
includes the Einstein-Yang-Mills (EYM) black hole with color index
$n$~\cite{EYMbh,KM,VG}  and Einstein-Born-Infeld (EBI) black hole
with Born-Infeld (BI) parameter $b^2$~\cite{SGP,Demianski}. The
difference is that the color charge $n$ is discrete, while the BI
parameter $b^2$ is continuous. However, a difficult issue has been
to analyze these hairy black holes completely.

Recently, Hod revisited  the EYM theory where hairy black hole was
found as the EYM black hole~\cite{Hod}. The hair extends beyond the
photon-sphere of the  black hole spacetimes. He argued that the
region beyond the photon-sphere involves more than half of the total
hair's mass.  It is well known   that the nonlinear character of the
YM fields plays a crucial role in constructing the hairy black
hole~\cite{Nunez}. More clearly, the nonlinearity of the YM fields
is  an essential tool in binding together the hair in two regions
such that  the near-horizon hair between event horizon and
photon-sphere does not collapse into the black hole, whereas the
far-region hair between photon-sphere and infinity does not escape
to infinity. Finally, he has proposed the hairy mass bound of
$m^+_{\rm hair}/m^-_{\rm hair} \ge1$, where $m^+_{\rm hair}(m^-_{\rm
hair}$) denotes the hair mass difference of far-region
(near-horizon) and photon-sphere. Here $m_{{\rm hair}}$ is the mass
of the hair that resides outside the horizon, i.e, $m_{{\rm
hair}}=m^+_{\rm hair}+m^-_{\rm hair}$.

On the other hand, a  charged  black hole was obtained from Einstein
gravity minimally coupled to nonlinear electro-magnetics,  which is
known as the EBI black hole. It is regarded as a nonlinear
generalization of the Reissner-Nordstr\"om (RN) black
hole~\cite{SGP,Demianski}.  At infinity, the EBI black hole with
$(M,Q,b^2)$ is indistinguishable from the RN black hole $(M,Q)$,
which implies that $b^2$ is considered as a free parameter like
color index $n$ for the EYM black hole. Indeed, Breton has applied
the isolated horizon formalism to the EBI black hole and, he has
shown clearly  that the BI parameter $b^2$ plays the same role of
color index $n$ for the EYM black hole~\cite{Breton}, although $n$
is the parameter of the solution while $b^2$ is the parameter of the
theory. Here, the BI parameter $b^2$ provides two limiting cases:
the limit of $b^2\to \infty$ corresponds to the RN black hole, while
in the limit of $b^2\to 0$ (or $Q=0$), one arrives at the
Schwarzschild black hole without charge $Q$.
\begin{figure*}[t!]
   \centering
   \includegraphics{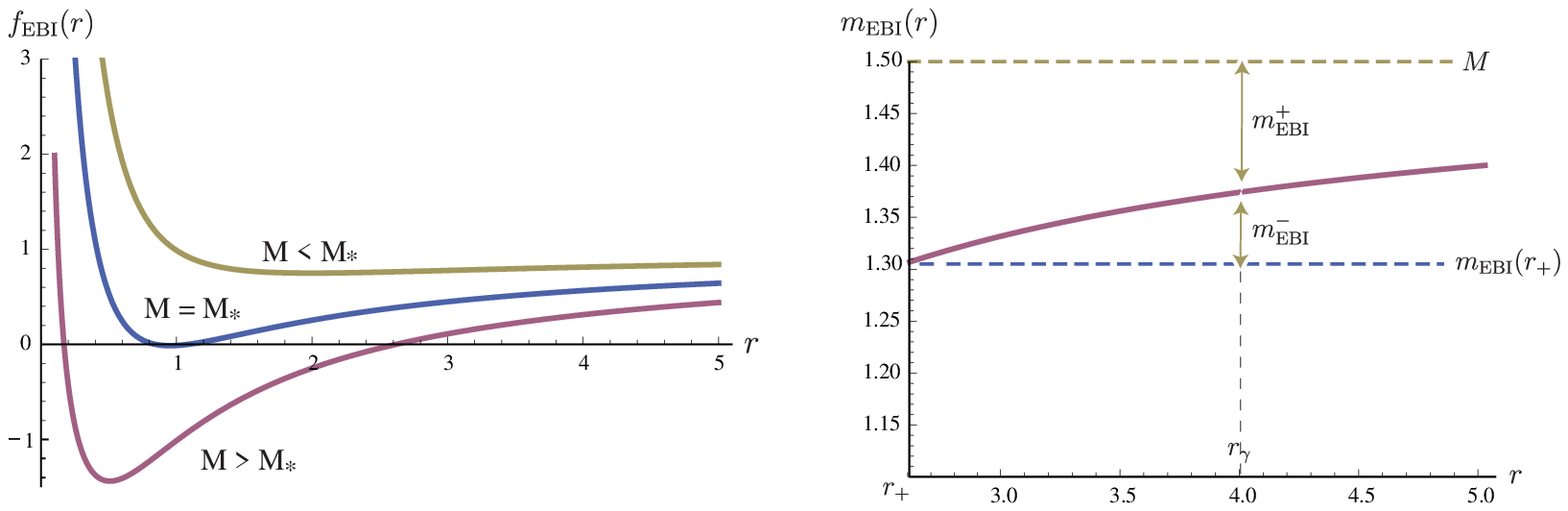}
\caption{(left panel) Metric function $f_{\rm EBI}(r)$ as a function
of $r$ with $b=2$, $Q_g=1$. For $M=M_{*}=0.99$ the degenerate
horizon is located at $r_{*}=0.9724$, while for $M>M_{*}$, the black
hole appears with the inner horizon $r_{-}=0.1696$ and the outer
horizon
$r_{+}=2.6181$. For $M<M_{*}$, there is no black hole.\\
\hspace*{3.2em}(right panel) Mass function $m_{{\rm EBI}}(r)$ as a
function of $r$ for $M=1.5,Q_g=1,b=2,r_+=2.61,r_\gamma=4$. In region
beyond the photon-sphere,  $m_{{\rm EBI}}^{+}$ involves more than
half of the total hair's mass.}
\end{figure*}

In this work, we study the EBI theory where hairy black hole was
found as bound state of  EBI black hole and EBIon  by making use of
Hod's idea. A key motivation is that the solution of EBI black hole
was known to take an analytic form unlike the EYM black hole in
which the solution was found by numerical way. Hence, it may provide
a good bed for testing the hairy mass bound.

We first introduce Einstein gravity coupled with nonlinear
electromagnetics known as the EBI action~\cite{Breton}
\begin{equation}\label{action}
S_{\rm EBI}=\int d^4x \sqrt{-g}\Bigg[\frac{R}{16\pi
G}+L(P,\tilde{Q})\Bigg],
\end{equation}
where the Born-Infeld Lagrangian is given by
\begin{equation}
L(P,\tilde{Q})=-\frac{P^{\mu\nu}F_{\mu\nu}}{2}+K(P,\tilde{Q})
\end{equation}
with the structural function  \be K(P,\tilde{Q})=
b^2\Bigg(1-\sqrt{1-\frac{2P}{b^2}+\frac{\tilde{Q}^2}{b^4}}\Bigg).
\ee In these expressions, $F_{\mu\nu}$ is defined by
$F_{\mu\nu}=\nabla_{\mu}A_{\nu}-\nabla_{\nu}A_{\mu}$ with a vector
potential $A_{\mu}$ and $P^{\mu\nu},~P,~\tilde{Q}$ are given by
\begin{eqnarray}
P^{\mu\nu}=2\frac{\partial L}{\partial F_{\mu\nu}},~~
P=\frac{1}{4}P_{\mu\nu}P^{\mu\nu},
~~\tilde{Q}=\frac{1}{4}P_{\mu\nu}\tilde{P}^{\mu\nu},
\end{eqnarray}
where $\tilde{P}^{\mu\nu}$ denotes the dual tensor of $P^{\mu\nu}$,
defined by
$\tilde{P}^{\mu\nu}=\epsilon^{\mu\nu\rho\sigma}P_{\rho\sigma}/2$.
Also, the coupling constant $b^2$ represents the BI parameter.
 Now we introduce the line
element with the metric function $f_{\rm EBI}(r)$ as follows:
\begin{equation} \label{metricBI}
ds_{\rm EBI}^2 = - f_{\rm EBI}(r) dt^2 + \frac{ dr^2}{f_{\rm
EBI}(r)} + r^2 \Big( d\theta^2+\sin^2\theta d\phi^2\Big).
\end{equation}
Choosing  a spherically symmetric background (\ref{metricBI}), the
electrically (magnetically) charged solution is obtained by  taking
\begin{equation}
F_{10}  =
\frac{Q}{\sqrt{r^4+\frac{Q^2}{b^2}}},~~P_{10}=\frac{Q}{r^2}.
\end{equation}

In this work, we consider the magnetically charged case only (from
here on we put $Q=Q_{g}$). The EBI  black hole solution can be
written as [see Fig.1 (left panel)]
\begin{eqnarray}
f_{\rm EBI}(r) &=& 1 - \frac{2M}{r}+ \frac{2 b^2 r^2}{3} \left( 1 -
\sqrt{ 1 + \frac{Q_{g}^2}{b^2r^4 }}
 \right)\nonumber\\&&\hspace*{8em}+\frac{ 4 Q_{g}^2}{ 3 r}G(r),  \label{fbi}\\
 G'(r)&=&-\frac{1}{\sqrt{ r^4 + \frac{Q_{g}^2}{b^2}}},
\end{eqnarray}
where $G'(r)$ denotes the derivative of $G(r)$ with respect to its
argument.
 $G$ takes the form \be\label{G1}
G(r)=\int^\infty_r\frac{ds}{\sqrt{s^4+\frac{Q_{g}^2}{b^2}}}=\frac{1}{r}F
\Big[\frac{1}{4},\frac{1}{2},\frac{5}{4};-\frac{Q_{g}^2}{b^2r^4}\Big],
\ee where $F$ is the hypergeometric function. In the presence of a
negative cosmological constant and  an electric charge, its solution
and thermodynamics were investigated  in~\cite{dey,Fernando,mkp}.
 On the other hand,
for the soliton solution, it takes the form
 \be
G(r)=-\int^r_0\frac{ds}{\sqrt{s^4+\frac{Q_{g}^2}{b^2}}}. \ee For
large $r$, the expansion of $f_{\rm EBI}$ with $G(r)$
[Eq.(\ref{G1})] is given approximately by
\begin{eqnarray}
\label{bis} f_{\rm EBI}(r)&\simeq&
1-\frac{2M}{r}+\frac{Q_{g}^2}{r^2}\nonumber\\
&&\hspace*{-5em}+\Bigg(-\frac{Q_g^4}{20b^2r^6}
+\frac{Q_g^6}{72b^4r^{10}}-\frac{5Q_g^8}{832b^6r^{14}}+\cdots\Bigg).
\end{eqnarray}
We would like to mention  two limiting cases as guided black holes
to understand the EBI black hole.  In the limit of  $Q_g \rightarrow
0$, the metric function reduces to the Schwarzschild case, while in
the limit of $b \rightarrow \infty$ and  $Q_g \neq 0$, the metric
function reduces to the RN case.

The relation between the ADM mass $M$ and the horizon radius $r_+$
is obtained by the condition of $f_{\rm EBI}(r_+)=0$~\cite{Breton}
\begin{eqnarray} \label{aas1}
M(r_+,Q_g,b) &=& \frac{r_{+}}{2} + \frac{ b^2 r_{+}^3}{3} \left( 1 -
\sqrt{ 1 + \frac{Q_{g}^2}{ b^2 r_+^4}}
\right)\nonumber\\
&&+\frac{2Q_{g}^2}{3r_+}F
\Big[\frac{1}{4},\frac{1}{2},\frac{5}{4};-\frac{Q_{g}^2}{b^2r_+^4}\Big].
\end{eqnarray}
This is possible because ``$M$-term" includes as  a single one in
the metric function $f_{\rm EBI}$.


Recently, Hod proposed that in the presence of a hairy black hole, a
non-trivial configuration of the hair extends above the null
circular geodesic (photon-sphere) of the black hole spacetimes.
According to Hod's approach \cite{Hod},  the radius $r_{\gamma}$ of
the photon-sphere in the hairy EBI black hole is determined by the
relation
\begin{eqnarray}\label{eqp}
&&2f_{\rm EBI}(r)-rf_{\rm EBI}^{\prime}(r)\Big{|}_{r=r_{\gamma}}=0\nn\\
&\to& 3r_\gamma\left(M-\frac{r_\gamma}{3}\right)=2Q_{g}^2 F
\Big[\frac{1}{4},\frac{1}{2},\frac{5}{4};-\frac{Q_{g}^2}{b^2r_\gamma^4}\Big]
\end{eqnarray}
which implies that the radius $r_\gamma$ is determined  completely
for given $(M,~Q_g,~b^2)$ by numerical way.
 Importantly, the nonlinear property of the BI fields provides a
key mechanism  in binding together the hair in two regions:
$m^{+}_{\rm EBI}$ and $m^{-}_{\rm EBI}$ where
\begin{eqnarray}\label{mp}
m^{+}_{\rm EBI}\equiv M-m_{\rm EBI}(r_{\gamma})
\end{eqnarray}
is the hair  mass difference  between photon-sphere and infinity
and
\begin{eqnarray}\label{mm}
m^{-}_{\rm EBI}\equiv m_{\rm EBI}({r_{\gamma}})-m_{\rm EBI}(r_{+})
\end{eqnarray}
is the hair mass difference   between event horizon and
photon-sphere.  In (\ref{mp}) and (\ref{mm}), the mass function
$m_{\rm EBI}(r)$ defined by
\begin{equation}
f_{\rm EBI}(r)\equiv 1-\frac{2m_{\rm EBI}(r)}{r}
\end{equation}
takes the form
\begin{eqnarray} \label{massfun}
m_{\rm EBI}(r)&=&M-\frac{ b^2 r^3}{3} \left( 1 - \sqrt{ 1 +
\frac{Q_{g}^2}{ b^2 r^4}} \right)\nn\\
&&-\frac{2Q_{g}^2}{3r}F
\Big[\frac{1}{4},\frac{1}{2},\frac{5}{4};-\frac{Q_{g}^2}{b^2r^4}\Big].
\end{eqnarray}
 At the horizon, the mass function takes a simple form
\begin{equation}
m_{\rm EBI}(r_+)=\frac{r_+}{2}.
\end{equation}
Hence,  $m_{\rm EBI}^{+}$ and $m_{\rm EBI}^{-}$ are computed  to be
\begin{eqnarray}
m_{\rm EBI}^{+}&=&\frac{ b^2 r_{\gamma}^3}{3} \left( 1 - \sqrt{ 1 +
\frac{Q_{g}^2}{ b^2 r_{\gamma}^4}}
\right)\nn\\
&&\hspace*{3em}+\frac{2Q_{g}^2}{3r_{\gamma}}F
\Big[\frac{1}{4},\frac{1}{2},\frac{5}{4};-\frac{Q_{g}^2}{b^2r_{\gamma}^4}\Big],\\
m_{\rm EBI}^{-}&=&\frac{ b^2 }{3} \left\{r_{+}^3\left( 1 - \sqrt{ 1
+ \frac{Q_{g}^2}{ b^2 r_+^4}} \right)-r_{\gamma}^3\times
 \right.\nn\\
&&\hspace*{-3em}\left. \left( 1-\sqrt{ 1 + \frac{Q_{g}^2}{ b^2
r_{\gamma}^4}}
\right)\right\}+\frac{2Q_{g}^2}{3}\left\{\frac{1}{r_+}F
\Big[\frac{1}{4},\frac{1}{2},\frac{5}{4};-\frac{Q_{g}^2}{b^2r_{+}^4}\Big]\right.\nn\\
&&\hspace*{3em}\left. -\frac{1}{r_{\gamma}}F
\Big[\frac{1}{4},\frac{1}{2},\frac{5}{4};-\frac{Q_{g}^2}{b^2r_{\gamma}^4}\Big]
\right\}.
\end{eqnarray}
Using the above expressions,  the ratio of distributions of the hair
outside the horizon can be obtained as
\begin{eqnarray}\label{EBIm}
\frac{m_{\rm EBI}^{+}}{m_{\rm EBI}^{-}}=\frac{1}{\frac{ b^2 r_{+}^3
\left( 1 - \sqrt{ 1 + \frac{Q_{g}^2}{ b^2 r_+^4}}
\right)+\frac{2Q_{g}^2}{r_+}F
\Big[\frac{1}{4},\frac{1}{2},\frac{5}{4};-\frac{Q_{g}^2}{b^2r_{+}^4}\Big]}
{ b^2 r_{\gamma}^3 \left( 1 - \sqrt{ 1 + \frac{Q_{g}^2}{ b^2
r_{\gamma}^4}} \right)+\frac{2Q_{g}^2}{r_{\gamma}}F
\Big[\frac{1}{4},\frac{1}{2},\frac{5}{4};-\frac{Q_{g}^2}{b^2r_{\gamma}^4}\Big]}-1}
\end{eqnarray}
We would like to mention that in the limit of $b\to\infty$, the
above ratio reduces to the case for the RN black hole,
\begin{eqnarray}\label{RNm}
\frac{m_{\rm RN}^{+}}{m_{\rm
RN}^{-}}=\frac{1}{\frac{r^{\rm RN}_{\gamma}}{r^{\rm RN}_{+}}-1}\geq 1,
\end{eqnarray}
where  $r^{\rm RN}_{\gamma}=\left[3M+\sqrt{9M^2-8Q^2}\right]/2$ and
$r^{\rm RN}_{+}=M+\sqrt{M^2-Q^2}$. The equality is achieved for the
extremal black hole  with $Q=M$.  We note that the bound (\ref{RNm})
is valid for  a linear electromagnetic theory. However, Hod
conjectured that this bound  persists in the nonlinear EYM
theory\footnote{Hod has tested the bound (\ref{generalm}) for large
EYM black holes and confirmed that it satisfies  the conjectured
bound (\ref{generalm}) as $m_{\rm EYM}^+/m_{\rm EYM}^{-}=2.08$ for
$n=1$.}:
\begin{eqnarray}\label{generalm}
\frac{m_{\rm hair}^{+}}{m_{\rm hair}^{-}}\geq 1.
\end{eqnarray}
On the other hand,  it is interesting to check if the EBI black hole
satisfies the conjectured bound (\ref{generalm}). For this purpose,
we calculate  the  ratio (\ref{EBIm}) by setting\footnote{Note that
we have tested the bound of $m_{{\rm EBI}}^{-}>m_{{\rm EBI}}^{+}$
for $b=(0.1,~1,~10,~100,~1000)$.}
$M=1.5,~Q_g=1,~b=2,~r_+=2.61,~r_{\gamma}=4$ as [see Fig.1(right
panel)]
\begin{eqnarray}\label{value1}
\frac{m_{\rm EBI}^{+}}{m_{\rm EBI}^{-}}=1.89
\end{eqnarray}
which implies that the EBI hair outside the horizon satisfies the
conjectured bound (\ref{generalm}).\\
\hspace*{1em} In summary, when taking into account $m^+_{\rm EBI}$
and $m^-_{\rm EBI}$ in the EBI black hole, the mass ratio ($m^+_{\rm
EBI}/m^-_{\rm EBI}$) is given by (\ref{value1}), which implies that
the EBI hair outside the horizon respects Hod's conjectured bound
(\ref{generalm}). It is therefore found that the EBI black hole
gives another example of showing that the region above the
photon-sphere
contains more than half of the total hair's mass.\\
\\
\hspace*{1em} This work was supported by the National Research
Foundation of Korea (NRF) grant funded by the Korea government
(MEST) through the Center for Quantum Spacetime (CQUeST) of Sogang
University with Grant No.2005-0049409. Y. Myung was partly supported
by the National Research Foundation of Korea (NRF) grant funded by
the Korea government (MEST) (No.2011-0027293).

\end{document}